\documentclass[a4paper]{cas-dc}

\usepackage[authoryear]{natbib}
\usepackage{amsmath}
\usepackage{amssymb}
\usepackage{graphicx}
\usepackage{algorithm}
\usepackage{algorithmic}
\usepackage{amsmath}
\usepackage{bm}
\usepackage{booktabs, tabularx}
\usepackage[section]{placeins}  

\def\tsc#1{\csdef{#1}{\textsc{\lowercase{#1}}\xspace}}
\tsc{WGM}
\tsc{QE}
\tsc{EP}
\tsc{PMS}
\tsc{BEC}
\tsc{DE}

\begin{document}
\let\WriteBookmarks\relax
\def\floatpagepagefraction{1}
\def\textpagefraction{.001}




\title [mode = title]{Binding Agent ID: Unleashing the Power of AI Agents with accountability and credibility}                      
\author[1]{Zibin Lin}[style=chinese]
\ead{linaacc9595@gmail.com}

\author[1]{Shengli Zhang}[style=chinese]
\cormark[1]
\ead{zsl@szu.edu.cn}

\author[1]{Guofu Liao}[style=chinese]
\ead{liaoguofu2022@email.szu.edu.cn}

\author[2]{Dacheng Tao}[style=chinese]
\ead{dacheng.tao@gmail.com}

\author[1]{Taotao Wang}[style=chinese]
\ead{ttwang@szu.edu.cn}

\affiliation[1]{organization={Shenzhen University},
    city={Shenzhen},
    country={China}}

\affiliation[2]{organization={Nanyang Technological University},
    country={Singapore}}

\cortext[cor1]{Corresponding author}





\begin{abstract}
Autonomous AI agents lack traceable accountability mechanisms, creating a fundamental dilemma where systems must either operate as ``downgraded tools'' or risk real-world abuse. This vulnerability stems from the limitations of traditional key-based authentication, which guarantees neither the operator's physical identity nor the agent's code integrity. To bridge this gap, we propose BAID (Binding Agent ID), a comprehensive identity infrastructure establishing verifiable user-code binding. BAID integrates three orthogonal mechanisms: local binding via biometric authentication, decentralized on-chain identity management, and a novel zkVM-based Code-Level Authentication protocol. 
By leveraging recursive proofs to treat the program binary as the identity, this protocol provides cryptographic guarantees for operator identity, agent configuration integrity, 
and complete execution provenance, thereby effectively preventing unauthorized operation and code substitution. We implement and evaluate a complete prototype system, demonstrating the practical feasibility of blockchain-based identity management and zkVM-based authentication protocol.

\end{abstract}




\maketitle






\section{Introduction}
\label{sec:introduction}

Artificial Intelligence (AI) agents are evolving from mere tools into autonomous entities capable of independent perception,
reasoning, decision-making, and action execution~\cite{Xi2025RiseLLMAgentsSurvey, guo2024large}. These agents can decompose
natural language objectives into executable steps and dynamically adjust their plans by invoking external APIs and
tools based on environmental feedback~\cite{Wang2024}. This capability enables them to bridge digital and physical
boundaries, opening unprecedented opportunities for automation and intelligent assistance. However, their deployment
faces a fundamental challenge: the absence of traceable and accountable responsibility chains. Without proper identity
and accountability mechanisms, this limitation creates a critical dilemma where systems must either be ``downgraded to tools''
requiring step-by-step human confirmation, or risk real-world abuse including voice cloning fraud,
mass information manipulation, and governance conflicts~\cite{Verma2023a,Verma2023b,chan2023harms,raskar2025upgrade}. These abstract
accountability challenges manifest concretely when autonomous agents perform actions in the real world that cross organizational
boundaries and require clear attribution of responsibility.

\subsection{Motivating Example: E-commerce Procurement}

\begin{figure}
    \centering
    \includegraphics[width=1\columnwidth,height=0.55\textheight,keepaspectratio]{ 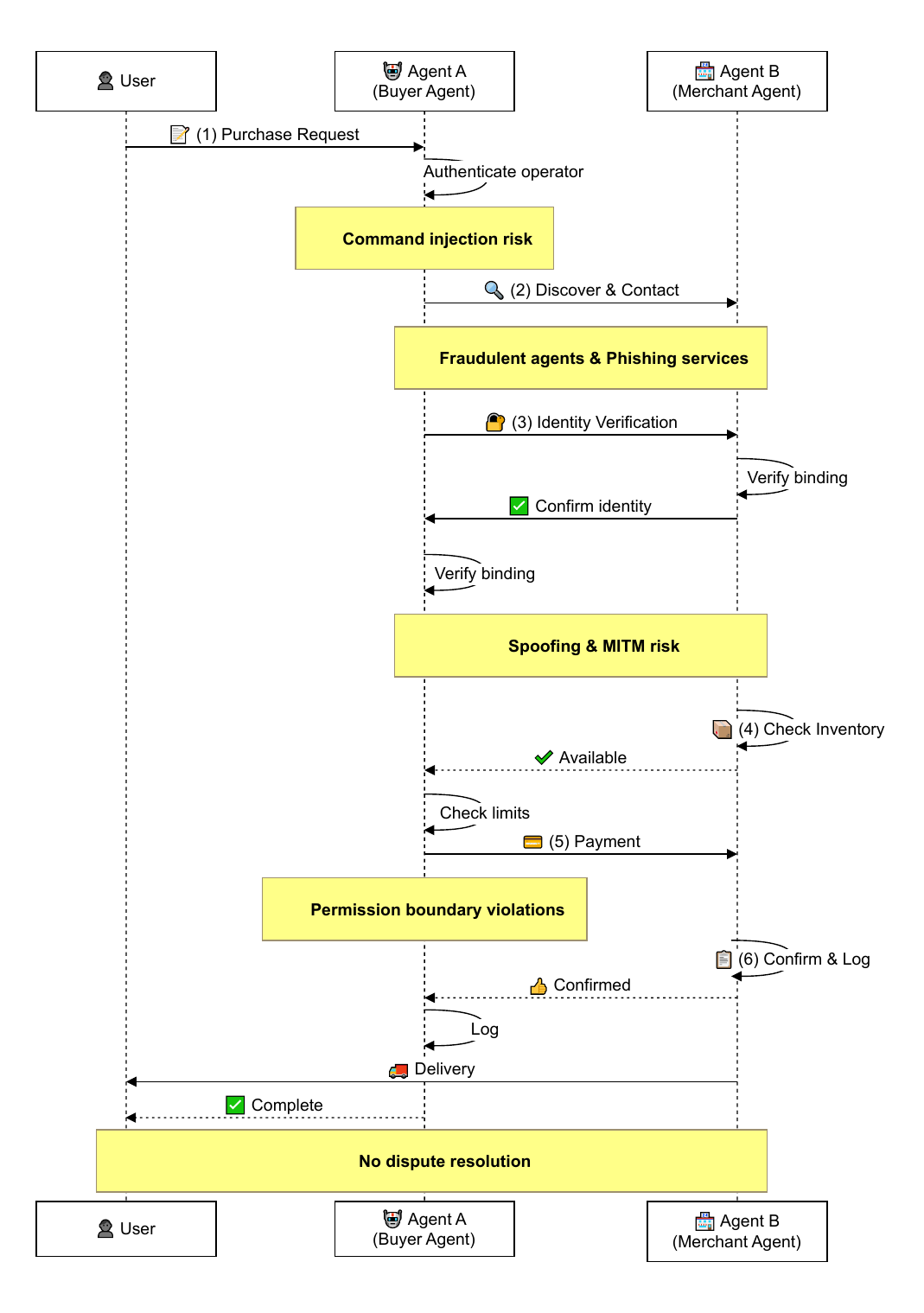}
    \vspace{-8mm}
    \caption{E-commerce laptop procurement workflow with security and accountability challenges across phases.}
    \label{fig:workflow}
\end{figure}

Consider a laptop procurement scenario where an AI agent (Agent~A) acting on behalf of a user must interact with a merchant's agent (Agent~B). The workflow showed in Fig.~\ref{fig:workflow} proceeds as follows:
(1)~Agent~A receives a purchase request from its user; (2)~Agent~A discovers and contacts merchant Agent~B;
(3)~both agents mutually verify identities; (4)~Agent~B checks inventory availability; (5)~Agent~A executes payment
within authorized limits; (6)~Agent~B confirms order and initiates delivery. This seemingly straightforward
transaction reveals fundamental security requirements across five critical security phases:

\begin{enumerate}[(1)]
\item \textit{User Identity Verification}: Agent~A must authenticate that its operator is the legitimate bound user.
Without this verification, \textbf{command injection attacks} can hijack the agent for unauthorized purchases.

\item \textit{Trusted Discovery}: Agent~A must discover and validate Agent~B through a secure directory system.
Without trusted discovery, Agent~A risks connecting to \textbf{fraudulent agents} or \textbf{phishing services}.

\item \textit{Mutual Authentication}: Both agents must bilaterally verify identities to establish trust.
Without robust mutual authentication, \textbf{identity spoofing} and \textbf{man-in-the-middle} attacks can
compromise transaction integrity.

\item \textit{Permission Control}: Agent~A must operate within strictly defined authorization policies following
the principle of least privilege. Inadequate controls enable \textbf{permission boundary violations} that exceed
spending limits or authorized scope.

\item \textit{Accountability Tracking}: The system must maintain comprehensive audit trails for responsibility
attribution and dispute resolution. Without auditable records, \textbf{no dispute resolution} to resolve disputes
or hold parties accountable.
\end{enumerate}

Without proper identity and accountability infrastructure, each phase presents significant security and
trust challenges that existing approaches cannot adequately address.

\subsection{The Gap: Autonomous Capabilities Without Identity Infrastructure}

Despite intensive research on system-level agent security---including goal alignment~\cite{Kirk2024PRISMAlignment, Huang2024CollectiveConstitutionalAI, OpenAI2024ModelSpec}, adversarial robustness~\cite{Zou2024CircuitBreakers, Anil2024ManyShotJailbreaking, Wallace2024InstructionHierarchy}, and multi-agent coordination~\cite{Yan2024GetItCooperating, Chen2024ComposableContracts, Tran2025MultiAgentCollaborationSurvey}---a fundamental infrastructure gap remains: the lack of verifiable identity and accountability. Existing approaches typically assume unified trust boundaries, overlooking critical questions of ``who is operating'' and ``who bears responsibility'' in open ecosystems.

Addressing this deficit requires a clear definition of responsibility. Theoretical models primarily distinguish between: (1)~the \textit{Autonomous Responsibility Model}, where agents independently bear liability (implying legal personhood); and (2)~the \textit{Binding Agent Model}, where agents act as accountable tools bound to specific human or organizational principals. We argue that the Binding Agent Model is the only practicable approach under current technological and legal constraints, as AI lacks legal subject status and agency law attributes actions to principals.

However, realizing the Binding Agent paradigm introduces a fundamental challenge stemming from the distinction between autonomous agents and standard software. Unlike passive tools, autonomous agents possess the capacity for independent decision-making and real-world interactions. This autonomy necessitates a stricter form of binding: the agent must not only serve its owner securely but also ensure that all consequences are irrefutably traceable to that principal. Consequently, mere local user authentication—which provides only service exclusivity—is insufficient. Instead, a \textit{third-party verifiable proof} is required to confirm that critical actions were explicitly authorized by the user. This obligation is absent in traditional software models, which typically rely on implicit local trust.

Emerging Agent Identity systems have laid meaningful foundations for naming and discovery: ANS~\cite{huang2025ans} provides DNS-like resolution, AgentFacts~\cite{nanda2025beyonddns} enables verifiable capability metadata, and works by Chan et al.~\cite{chan2024ids} and Raskar et al.~\cite{raskar2025upgrade} integrate identity markers into existing infrastructure. Recently, decentralized frameworks like LOKA~\cite{ranjan2025loka} and ERC-8004~\cite{erc8004} have further advanced this field by addressing ethical alignment and on-chain validation. However, while these advancements successfully establish ``agent-to-system'' trust—ensuring agents are locatable, valid, and compliant entities—they fall short of establishing ``human-to-agent'' liability binding. They lack the mechanism to cryptographically anchor agent actions to a liable human principal. By relying on traditional key-based authentication, they inherit the semantic gap of standard software security: keys prove possession but neither authenticate the operator's physical identity nor guarantee the execution logic's integrity. In open environments, third-party verifiers require cryptographic assurance that an agent's actions are authorized by its bound user—a mechanism absent in existing systems, leaving them vulnerable to code substitution and accountability evasion. To address this, we innovatively employ Zero-Knowledge Proofs (ZKP) to generate verifiable proofs of the authentication process itself, effectively bridging the gap between local authorization and global accountability.
\subsection{Our Contributions}

To address these challenges and implement the Binding Agent responsibility paradigm, we propose the BAID (Binding Agent ID) framework. Our specific contributions are as follows:

\begin{itemize}
    \item \textbf{We propose the BAID framework}, a comprehensive identity infrastructure that integrates local biometric binding, decentralized on-chain identity management, and verifiable agent authentication. This framework provides a unified solution for identity verification, trusted discovery, and accountability tracking, ensuring ``the correct agent is operated by the binding user executing committed code''.
    
    \item \textbf{We introduce a zkVM-based agent authentication protocol} to establish verifiable user-code binding. By leveraging code-level authentication and recursive zero-knowledge proofs, this mechanism cryptographically binds the agent's execution to its registered identity and authorized user, effectively preventing code substitution attacks and ensuring execution integrity without revealing sensitive operational details.
    
    \item \textbf{We implement a complete prototype and provide extensive evaluation}. We develop the full system architecture including the identity management smart contracts, and zkVM verification circuits. Our experimental results demonstrate the practical feasibility of the system, with gas-efficient on-chain operations and millisecond-level verification latency suitable for real-world deployment.
\end{itemize}

The rest of this paper is organized as follows. Section~\ref{sec:preliminaries} introduces the technical 
preliminaries. Section~\ref{sec:overview} presents the BAID system architecture and technical overviews. Section~\ref{sec:design} describes the detailed technical 
design of the binding mechanisms. Section~\ref{sec:evaluation} evaluates the system's security properties, performance 
characteristics, and deployment feasibility across multiple dimensions. Section~\ref{sec:related} surveys related work. Section~\ref{sec:conclusion} concludes the paper.

\section{Technical Preliminaries} \label{sec:preliminaries}
This section introduces the fundamental technologies underlying our BAID framework.

\subsection{Blockchain}

Blockchain represents a \textit{decentralized}, append-only ledger characterized by three fundamental properties essential for agent identity systems: \textit{decentralization}, \textit{immutability}, and \textit{public transparency} \cite{nakamoto2008bitcoin}.
Through distributed consensus protocols (e.g., Proof-of-Work, Proof-of-Stake), all state transitions are cryptographically linked and verifiable by any participant, eliminating reliance on centralized authorities while ensuring that identity credential registrations, delegations, and revocations remain fully auditable.
Smart contracts—deterministic programs deployed on blockchains such as Ethereum \cite{buterin2014ethereum}—extend this foundation by enabling Turing-complete logic execution within the Ethereum Virtual Machine (EVM).
Upon transaction triggers, contract code and state modifications are validated by all consensus participants, guaranteeing that identity lifecycle operations (issuance, verification, revocation) exhibit both \textit{tamper-evidence} and \textit{public verifiability}.
These properties collectively establish blockchain as a trust anchor for credential provenance and delegation chains in decentralized agent identity frameworks.

\subsection{Zero-Knowledge Virtual Machine}

The emergence of blockchain technology, distributed ledgers, and multi-party computation has created an increasing demand for efficient, scalable, and composable verification of arbitrary computation results while preserving input privacy. Zero-Knowledge Proof (ZKP) technology establishes the cryptographic foundation for achieving both computational integrity and privacy \cite{GoldwasserEtAl85}. Nevertheless, traditional approaches based on circuit-level or arithmetic descriptions, such as Rank-1 Constraint System and Quadratic Arithmetic Programs, present significant challenges for developers and require substantial effort to integrate with existing software ecosystems.

Zero-Knowledge Virtual Machine (zkVM) addresses this challenge by bridging the semantic gap between general-purpose programs and provable instances. By introducing an abstraction layer that encompasses an instruction set architecture (typically resembling traditional CPU designs or optimized RISC-style instructions) and a corresponding execution environment, zkVM automatically transforms program execution into verifiable constraints and proof generation procedures. This transformation enables proof-carrying execution while maintaining high degrees of development reusability and ecosystem compatibility.

Formally, let $\lambda$ denote the security parameter and let $\mathbb{F}_p$ be the base field where $p$ is a large prime. We model zkVM as a five-tuple:
\[ \Pi = (\text{Setup}, \text{CommitProg}, \text{Compile}, \text{Prove}, \text{Verify}) \]
consisting of the following algorithms:

\begin{enumerate}
\item $(pp, vk) \leftarrow \text{Setup}(1^\lambda, param)$: A setup algorithm that generates public parameters and verification key, where $param$ encompasses system parameters including the maximum step bound $T_{max}$, field specifications, and commitment scheme parameters.

\item $C_P \leftarrow \text{CommitProg}(P)$: A commitment algorithm that produces a cryptographic commitment $C_P$ to the program code $P$ using either hash-based, vector, or KZG commitment schemes.

\item $(\text{ArithDesc}, \text{MapIO}) \leftarrow \text{Compile}(pp, P)$: A compilation algorithm that performs static analysis and transforms the instruction set into constraint templates, yielding both arithmetic descriptions (constraint skeletons) and I/O mapping specifications. These outputs serve as static intermediates for subsequent proof generation, enabling the transformation of program logic into verifiable constraints.

\item $\pi \leftarrow \text{Prove}(pp, C_P, \text{ArithDesc}, \text{MapIO}, x_{pub}, x_{prv}, y)$: A proving algorithm that:
\begin{itemize}
    \item Constructs the execution trace
    \[\text{Trace}(P, x_{pub}, x_{prv}) = (s_0,\ldots,s_T),\]
    utilizing $\text{MapIO}$ to map public inputs $x_{pub}$ and private inputs $x_{prv}$ into the initial state and ensure output $y$ consistency in the final state.
    \item Generates the constraint representation by instantiating $\text{ArithDesc}$'s constraint templates with the execution trace (utilizing R1CS, AIR, or PLONKish frameworks), incorporating $\text{MapIO}$ for I/O constraints.
    \item Executes the underlying proof system to produce the proof $\pi$.
\end{itemize}

\item $b \leftarrow \text{Verify}(vk, C_P, x_{pub}, y, \pi)$: A verification algorithm that outputs a binary decision: accept (1) or reject (0). The verification implicitly relies on the constraint structure derived from $\text{ArithDesc}$ and $\text{MapIO}$ (embedded in $vk$ or $C_P$), ensuring the proof $\pi$ attests to correct program execution and I/O mapping without revealing private inputs.
\end{enumerate}

In the BAID framework, zkVM enables agents to generate \textit{verifiable execution proofs} demonstrating identity authenticity and operational compliance without revealing sensitive configurations or proprietary code.
These proofs cryptographically establish that agents operate under certified configurations, authenticated operators, and committed program logic, thereby achieving accountability through \textit{computational verification} rather than \textit{data disclosure}.

\subsection{Verifiable Credentials}
\label{sec:verifiable_credentials}

Verifiable Credentials (VC) and Decentralized Identifiers (DID) constitute the foundational standards for \textit{decentralized}, \textit{privacy-preserving}, and \textit{interoperable} digital identity systems \cite{w3c2025vc,w3c2022did}.
The trust model involves four key entities: \textit{Issuers} make cryptographically signed claims about subjects; \textit{Holders} store VCs and generate selective-disclosure Verifiable Presentations (VPs); \textit{Verifiers} authenticate credentials by resolving issuer and holder public keys through \textit{Registry Authorities}, which maintain immutable trust anchors recording entity identities, associated keys, and revocation lists.
This architecture enables privacy-preserving identity verification: holders can prove specific attributes (e.g., ``age $\geq$ 18'', ``KYC approved'') without revealing underlying personal data.

Integration with zero-knowledge proofs strengthens privacy guarantees \cite{pauwels2021zkkyc}.
In zkKYC scenarios, approved institutions (Issuers) generate signed VCs with cryptographic field commitments using privacy-preserving signature schemes such as BBS+ or CL-Signature.
Holders then generate Selective-Disclosure Proofs through local ZKP circuits, cryptographically demonstrating compliance with regulatory requirements (e.g., ``not sanctioned'', ``credential unexpired'', ``age $\geq$ 18'') while concealing sensitive personal identifiers.
Verifiers validate only the VC signature chain and ZKP correctness, confirming regulatory compliance without accessing underlying personal data.
This cryptographic separation of \textit{authentication} (proving credential validity) from \textit{disclosure} (revealing specific attributes) establishes VCs as the privacy-preserving credential substrate for BAID framework.

\section{BAID System Overview}\label{sec:overview}

\begin{figure*}
    \centering
    \includegraphics[width=0.65\textwidth]{  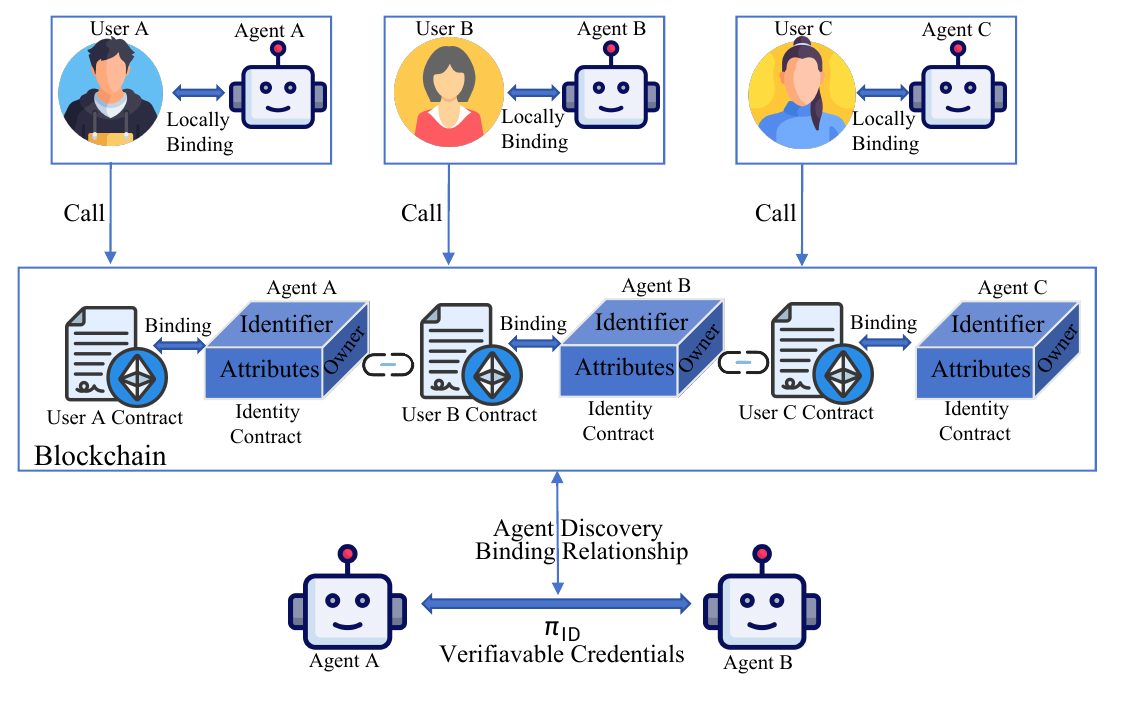}
    \caption{BAID system architecture: 1. Local User-Agent Binding mechanism; 2. On-Chain Identity Management; 3. Agents Authentication and Authorization Protocol.}
    \label{fig:baid-architecture}
\end{figure*}

The BAID (Binding Agent ID) system establishes verifiable, traceable, and accountable digital identity chains for autonomous AI agents, addressing the fundamental challenge of agent trustworthiness and cross-system collaboration. As illustrated in Figure~\ref{fig:baid-architecture}, the system workflow proceeds through three integrated stages: (1)~users establish local biometric binding with agents through the Biometric Authentication Module (BAM), anchoring agent identity to its human owners; (2)~users register Agent Identifiers on-chain via blockchain smart contracts, creating publicly queryable user-agent binding declarations that enable trusted agent discovery; (3)~agents authenticate to third-party verifiers through zkVM execution proofs combined with Verifiable Credentials, establishing cryptographically verified identity and permission scope for secure collaboration.
To implement this workflow, the BAID system architecture comprises three principal technical modules:

\textbf{(1) Local User-Agent Binding.} To support local binding and subsequent Agent ID management, we design an extended agent framework that enhances the Profile module and introduces an Identity Management module supporting biometric authentication and cryptographic operations. Users establish local binding by storing their biometric identity features in the agent's Profile security configurations file. Subsequently, the agent continuously validates operators against stored biometric templates through the Biometric Authentication Module (BAM), ensuring only the legitimate owner can activate the agent---this corresponds to Phase~2 (Operator Biometric Authentication) in the verification pipeline, thereby implementing the security property ``agent serves only its owner.''

\textbf{(2) On-Chain Identity Management.} Building upon this local foundation, leveraging blockchain's transparency, immutability, and decentralization, BAID constructs a globally unified trusted identity system supporting user/agent registration, on-chain binding, and agent discovery. Users complete real-person verification through zkKYC (see Section~\ref{sec:verifiable_credentials}) without revealing identity privacy, creating User Identity Contracts that establish legal entity attributes. Based on the User Identity Contract, users can subsequently create Agent Identity Contracts for locally-bound agents, making binding relationships publicly queryable. The Agent Identity Contract records agent attributes (user identity contract address, Agent Identifier, capabilities, roles) and communication metadata (protocols, endpoints, domains), thereby enabling agent discovery functionality that allows other agents to rapidly identify and locate collaboration channels. The on-chain Agent Identifier serves as the canonical reference for Phase~1 (Agent Configuration Integrity) verification, ensuring local configurations match blockchain-anchored versions.

\textbf{(3) Agents Authentication and Authorization Protocol.} With the identity infrastructure established, the authentication mechanism enables verifiers to validate agent identities and permission boundaries, establishing trust channels. As shown in Figure~\ref{fig:baid-architecture}, agents initially establish communication through the blockchain identity system's discovery functionality, but further collaboration requires mutual identity and permission verification. As software entities, agents' program commitments serve as their ``biometric identity features.'' We use program commitments as Agent Identifiers and employ zkVM proofs to verify consistency between currently executing programs and Agent Identifiers---this authentication mechanism transcends traditional key-based authentication limitations by simultaneously verifying the operator's physical identity and the agent's computational identity, effectively preventing both unauthorized operation and malicious program impersonation. Moreover, BAID implements verifiable permission authorization through Verifiable Credentials (VC), enabling users to perform fine-grained authorization combining agent attributes and target task contexts. Consequently, during agent authentication, agents must provide two proofs to verifiers: zkVM execution proof and user permission authorization proof (VC).

\section{BAID System Designs}\label{sec:design}

Having outlined the overall BAID architecture, we now provide detailed system designs for the framework's three core pillars. This section first presents the local binding infrastructure that establishes the foundational trust relationship between agents and their human owners (Section 4.1). It then details the blockchain-based identity registration and discovery mechanisms that enable global agent interoperability (Section 4.2). Finally, it describes the zkVM-based agent identity authentication protocol that facilitates verifiable and privacy-preserving interactions through code-level proofs (Section 4.3).

\subsection{Local User-Agent Binding via Biometric Authentication}

The local binding mechanism ensures that agents serve exclusively their designated owners by embedding user biometric identity directly into agent configurations. This mechanism is implemented through an enhanced agent architecture that integrates biometric authentication capabilities within the Identity Management module.

\subsubsection{Agent Architecture}

Contemporary agent frameworks typically consist of Profile (configuration), Memory (knowledge retention), Planning (decision-making), and Action (task execution) modules \cite{wang2024survey}. However, to support autonomous identity management and cryptographic operations, we extend this architecture with an Identity Management module and augment the Profile module with identity-binding capabilities.

\textbf{Profile Module.} Beyond recording agent attributes (capabilities, behavioral constraints, operational parameters), the Configuration Document within the Profile module stores the owner's Human Identifier and Biometric Authentication Credential. These identity anchors provide the foundation for deep binding and local user authentication mechanisms.

\textbf{Identity Module.} This module manages all Agent ID-related operations, consisting of two subcomponents: (1) Cryptographic module
  (e.g., zkVM) for generating authentication proofs and managing Verifiable Credentials; (2) BAM integration for real-time operator identity verification. The module interfaces with both local biometric sensors and blockchain identity contracts to enforce the security property ``agent serves only its owner.''

\begin{figure*}
    \centering
    \includegraphics[width=0.8\textwidth]{  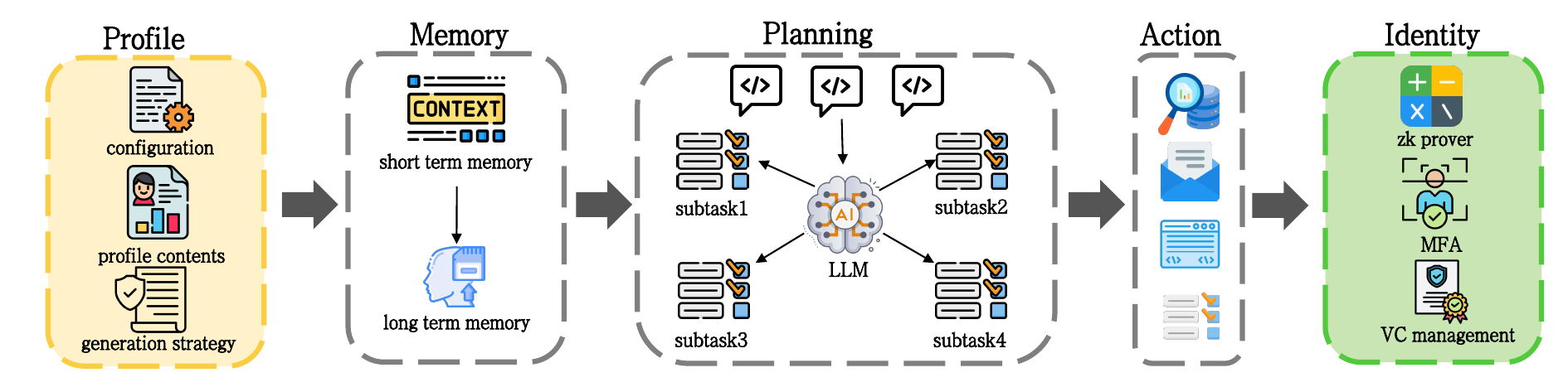}
    \caption{Extended agent architecture with Identity module: integrating biometric authentication, cryptographic operations, and identity binding capabilities into the standard agent framework (Profile, Memory, Planning, Action).}
    \label{fig:agent-architecture}
\end{figure*}

Figure~\ref{fig:agent-architecture} illustrates the enhanced agent architecture with identity management capabilities integrated into the standard agent framework.

\subsubsection{Biometric Template Registration and Verification}

The local binding mechanism establishes a cryptographically secure association between user biometric identity and agent configuration, operating through two distinct phases: registration and verification.

\begin{figure}
    \centering
    \includegraphics[width=1\columnwidth]{  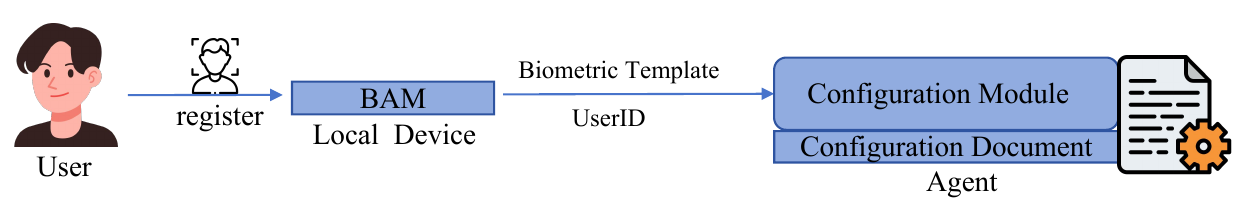}
    \vspace{-3mm}
    \caption{Local binding registration workflow: biometric template generation and configuration document binding.}
    \label{fig:local-binding}
\end{figure}

\textbf{Registration Phase.} During the initial binding establishment, the system performs the following steps as illustrated in Figure~\ref{fig:local-binding}:

\textbf{Step1: Biometric Capture and Template Generation.} The user invokes the device's local Biometric Authentication Module (BAM), which captures biometric data via facial recognition camera, fingerprint sensor, or iris scanner. BAM applies feature extraction algorithms to generate a Biometric Template---a compact mathematical representation of distinctive biological traits.

\textbf{Step2: Configuration Binding.} The system writes both the User Identifier and Biometric Template (designated as Registered Template) into the agent's Configuration Document within the Profile module, creating an immutable binding record that anchors the agent to its designated owner.

\textbf{Verification Phase.} Once bound, the Identity Module enforces continuous operator verification before executing sensitive operations. BAM captures fresh biometric samples, generates temporary templates, and performs similarity matching against the registered template to confirm identity. If verification fails, the agent immediately terminates or refuses service, ensuring that only the legitimate bound user can control agent actions.

\subsection{On-Chain Identity Management}

Building upon the local binding infrastructure, we construct a globally unified, decentralized on-chain identity system based on blockchain technology to support trusted identity management and secure collaboration across the agent ecosystem. This identity system not only strengthens user sovereignty over agents but also provides critical infrastructure for building secure, trustworthy, multi-party collaborative agent networks.

\subsubsection{Identity Management Smart Contract Framework}

The decentralized identity system comprises three primary architectural components:

\textbf{Entrypoint Contract.} Drawing inspiration from the ERC-4337 account abstraction protocol, we employ an Entrypoint Contract as the user-contract interaction gateway. The ERC-4337 protocol enables user accounts to execute complex programmable logic including gas sponsorship for agents (users pay transaction fees on behalf of their agents), delegated payment authorization (granting agents specific payment permissions within defined limits), and subordinate account binding (establishing agent identity accounts as dependent sub-accounts under user accounts). The Entrypoint Contract leverages these capabilities to verify and execute user transactions while invoking target identity contracts, enabling flexible identity management operations without requiring agents to hold native blockchain tokens for transaction fees.

\textbf{Identity Contracts.} The User Identity Contract and Agent Identity Contract serve as the respective account contracts for users and agents within the identity system, both instantiated by the Entrypoint using predefined templates. The User Identity Contract template contains variables including \texttt{Owner}, \texttt{userID}, and \texttt{Agent\_Binding\_List}, storing the user's public key, user identifier, and the list of $\mathsf{AgentID}$s bound to the user, respectively. The Agent Identity Contract template contains variables including \texttt{Owner\_User}, \texttt{OperationalStatus}, \texttt{Attribute}, and \texttt{AgentFactsURL}. These variables store: (1) the owner's contract address; (2) the agent's operational status (Running/Stopped/Deregistered); (3) agent attributes (including agent name, program commitment $C_P$, configuration hash, $\mathsf{userID}$, $\mathsf{AgentID}$, capabilities, and roles); and (4) the Uniform Resource Locator pointing to the agent's verifiable metadata (AgentFacts).

AgentFacts is a lightweight dynamic metadata document published by the agent, employing JSON-LD format with Verifiable Credential signatures to describe an AI agent's identity, capabilities, endpoints, routing, and trust status. It decouples stable index records from high-frequency changing information, supporting rapid updates, verifiable trust, and privacy-friendly resolution~\cite{nanda2025beyonddns}. The complete AgentFacts document supports multiple flexible storage options including agent self-hosting, IPFS decentralized storage, third-party hosting, CDN distribution, and storage within the BAID Agent Identity Contract itself. Only the \texttt{AgentFactsURL} needs to be published in the Agent Identity Contract to enable BAID's Agent Discovery functionality.

The identity system implements three core functions for lifecycle management, as detailed in Table~\ref{tab:core-functions}.

\begin{table}[htbp]
\caption{Core Functions for Agent Identity Lifecycle Management}
\label{tab:core-functions}
\centering
\small
\begin{tabularx}{\columnwidth}{lX}
\toprule
\textbf{Function} & \textbf{Description} \\
\midrule
\texttt{addAgent(...)} &
Creates Agent Identity Contract through Entrypoint, adds AgentID to \texttt{Agent\_Binding\_List}, establishing on-chain User-Agent binding for discovery and accountability. \\
\midrule
\texttt{updateAgent(...)} &
Modifies agent attributes after verifying caller authorization (\texttt{Owner\_User}), recording timestamps and versions for capability evolution tracking. \\
\midrule
\texttt{removeAgent(...)} &
Sets \texttt{OperationalStatus} to Deregistered, removes AgentID from binding list while preserving audit trails for compliance. \\
\bottomrule
\end{tabularx}
\end{table}

\subsubsection{On-Chain Identity Registration and Binding Workflow}

\begin{figure}
    \centering
    \includegraphics[width=1\columnwidth]{  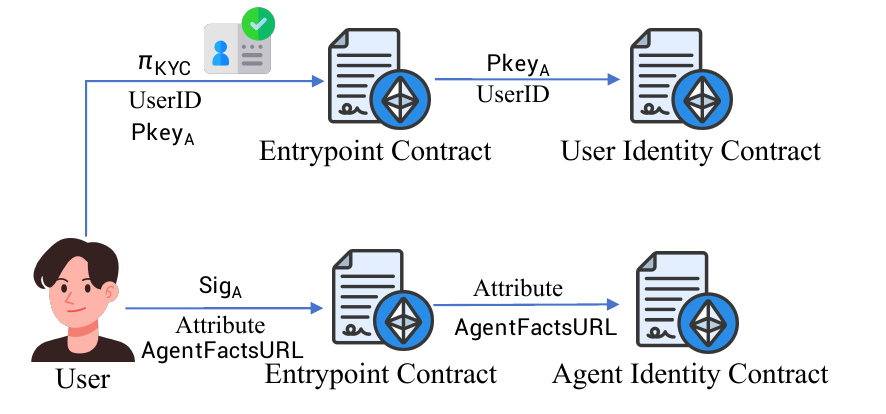}
    \caption{Registration and binding workflow showing UserID registration via zkKYC, AgentID generation, and on-chain AgentID registration.}
    \label{registration-binding}
\end{figure}

As illustrated in Figure~\ref{registration-binding}, the registration and binding workflow covers the complete agent lifecycle from user identity establishment to agent on-chain registration:

\textbf{Step1: UserID Registration.} Based on the principle of legal entity binding with optional anonymity, we leverage real-person zkKYC technology to confirm users' legal entity attributes and complete UserID registration without exposing users' real identities. After users pass zkKYC verification, the Entrypoint Contract creates a User Identity Contract and initializes the \texttt{userID} and \texttt{Owner} variables (\texttt{Owner = Pkey\_A}, where \texttt{Pkey\_A} is the user's blockchain account public key).

\textbf{Step2: AgentID Generation.} In BAID, the Agent Identifier follows the structure:
\begin{equation}
\label{eq:baid-structure}
\begin{split}
\mathsf{AgentID} = \text{agentid}:H(&\text{name} \| C_P \| \\
&H(\text{profile}) \| \text{userID} \| \text{others})
\end{split}
\end{equation}

where $C_P = \text{CommitProg}(P)$ is the program commitment,
$H(\text{profile})$ hashes the security configuration (user identity, policy constraints),
$\mathsf{userID}$ identifies the bound user, and $\text{name}$ provides human readability. The $\mathsf{AgentID}$ is generated locally by the user, preparing for subsequent on-chain identity registration. Table~\ref{tab:baid-example} presents an $\mathsf{AgentID}$ for a laptop retail sales advisor agent, demonstrating how these identity components are instantiated in practice.

\textbf{Step3: AgentID Registration and Binding.} The user generates variable data \texttt{Attribute}, \texttt{AgentFactsURL}, and a signature $\mathsf{Sig}_A$ using the owner's public key, then invokes the Entrypoint Contract to execute the User Identity Contract's \texttt{addAgent(AgentID, Name, Attribute, AgentFactsURL)} function (see Table~\ref{tab:core-functions}), adding the newly bound agent's information and creating the Agent Identity Contract. After signature verification passes, the Entrypoint Contract creates an Agent Identity Contract and initializes the \texttt{Owner\_User}, \texttt{Attribute}, and \texttt{AgentFactsURL} variables. The \texttt{Owner\_User} is set to the User Identity Contract's address, declaring the binding relationship between the user and agent and ownership of the Agent Identity Contract.

\begin{table}[htbp]
\caption{Example AgentID for Laptop Retail Sales Advisor Agent}
\label{tab:baid-example}
\centering
\small
\begin{tabular}{lp{0.6\columnwidth}}
\toprule
\textbf{Field} & \textbf{Value} \\
\midrule
name & \texttt{baid:agent:laptopretail:salesadvisor} \\
code\_hash & \texttt{sha256:2c1d7fa9c5b84d6f9a1e3d...} \\
profile\_hash & \texttt{sha256:8af4c61bd0b9e2a7d17f...} \\
UserID & \texttt{org:laptopretail:cn:sales-division} \\
Other & \texttt{``agent\_version'': ``1.0.0''} \\
AgentID & \texttt{agentid:5f04cf173a987be6a98e...} \\
\bottomrule
\end{tabular}
\end{table}

\subsection{zkVM-Based Agent Identity Authentication Protocol}
\label{sec:zkvm_auth}

Having established the local binding and on-chain registration mechanisms, we now present the zkVM-based agent identity authentication protocol that enables secure agent interactions with third-party verifiers through cryptographically verifiable identity and permission proofs. This protocol addresses the fundamental challenge of establishing trust when agents interact with external entities---including other agents, service providers, and regulatory authorities. Verifiers must authenticate both the agent's identity integrity and its delegated permission scope to establish a cryptographic trust foundation.

The authentication protocol comprises two complementary components: user-to-agent authorization through Verifiable Credentials, and agent identity authentication through zkVM proofs. The authorization mechanism enables users to delegate fine-grained permissions to agents, while the authentication mechanism enables agents to prove their identity and authorization scope to any third-party verifier without revealing sensitive information. Agent-to-agent interaction represents a primary application scenario, where mutual authentication enables trustworthy collaboration between autonomous systems.

\subsubsection{User-to-Agent Authorization}

Authorization establishes the permission delegation chain from users to agents, enabling agents to access user data, invoke resources, or act on behalf of users within explicitly defined boundaries. Our authorization mechanism features context-aware permission control that combines agent attributes with target task contexts, thereby preventing permission over-provisioning and protecting user data privacy.

\begin{figure}
    \centering
    \includegraphics[width=1\columnwidth]{  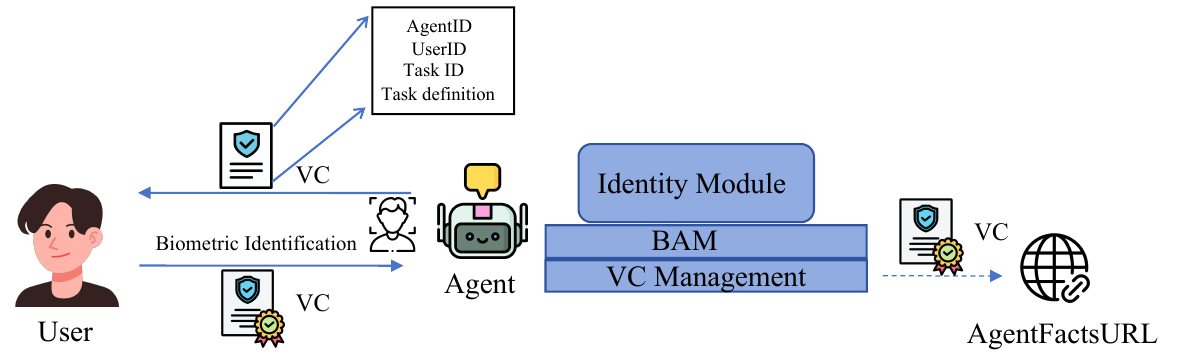}
    \caption{Authorization workflow showing VC generation, biometric authentication, and credential signing.}
    \label{fig:authorization-workflow}
\end{figure}

\textbf{Authorization Workflow.} As illustrated in Figure~\ref{fig:authorization-workflow}, the authorization process comprises three steps:

\textbf{Step1: Verifiable Credential Generation.} The agent's Identity Module generates a Verifiable Credential (VC) based on the agent's role, capabilities, and target task context. The VC contains: (1) Identity information (agent's $\mathsf{AgentID}$ and user's $\mathsf{UserID}$); (2) Task information (Task ID, Task definition); and (3) Permission information (security level, authorization scope, and credential validity period).

\textbf{Step2: Local User Authentication.} The Identity Module initiates local user authentication to verify that the current operator is the legitimate owner, following the biometric verification procedure described in Section~5.1.

\textbf{Step3: Credential Signing and Storage.} After passing local user authentication, the Identity Module signs the VC using the Owner's private key and attaches the signature to the VC. The signed VC is then stored in the agent's Identity Module. When necessary, the VC can be published to the AgentFactsURL, for example, VCs that can demonstrate the agent's capabilities or harmlessness.

\subsubsection{zkVM-Based Agent Identity Authentication}

Building upon the user-to-agent authorization foundation, the agent identity authentication protocol implements a dual-layer verifiable computation framework leveraging zkVM technology. This protocol addresses three fundamental technical challenges inherent to autonomous agent systems to establish cryptographic guarantees for agent identity integrity, operator legitimacy, and execution provenance.

\textbf{Challenge 1: Identity-Execution Binding.} Traditional key-based authentication is insufficient for autonomous agents, as possession of a signing key guarantees neither the integrity of the executing code nor the authenticity of the operator. An adversary with a compromised key could therefore execute malicious logic or impersonate the legitimate operator while masquerading as the authorized agent. To resolve this semantic gap, we introduce \textit{Code-Level Authentication} using zkVM. By establishing the program binary as the agent's identity and generating cryptographic proofs of execution, we bind the agent's identity directly to its computational behavior and operator authorization, ensuring that only the committed code executed by the bound user can generate valid authentication proofs.

\textbf{Challenge 2: Execution Continuity.} Agents operate through multi-turn interactions where state transitions must be sequentially valid. Standard verifiable computation proofs verify individual steps but lack mechanisms to enforce the correct ordering of a sequence, leaving systems vulnerable to replay or reordering attacks. We address this by leveraging zkVM's \textit{Recursive Verification} capability. Each execution step generates a proof that recursively verifies the validity of the previous step's proof, constructing an unbreakable cryptographic chain that guarantees the integrity and continuity of the entire interaction history.

\textbf{Challenge 3: Data Provenance.} Agents frequently interact with external environments (e.g., calling LLM APIs or accessing web data), but the zkVM cannot inherently verify the authenticity of external data. This creates a trust gap where an attacker could feed fabricated data to the agent. We bridge this gap by integrating \textit{zkTLS} (Zero-Knowledge Transport Layer Security). This allows the agent to generate proofs attesting that specific data originated from a trusted TLS-authenticated server (e.g., OpenAI), ensuring end-to-end provenance from external data sources to internal computation.

To systematically resolve these challenges, we propose a \textbf{Dual-Layer Verification Framework} (detailed technical specifications including mathematical proofs, security properties, and protocol formalization are provided in Appendix A). The protocol orchestrates verification through two integrated layers:

\textbf{Layer 1: Three-Phase Verifiable Computation Pipeline.} This layer implements end-to-end cryptographic guarantees through recursive proof algorithm across three sequential phases:

\textit{Phase 1---Operator Biometric Authentication.} To address the unique security requirements of autonomous agents, this phase validates operator legitimacy through privacy-preserving facial recognition. This mechanism fulfills two critical security objectives: it strictly enforces that the agent serves \textit{exclusively its designated owner}, and simultaneously generates a \textit{third-party verifiable proof} of this authorization—a mandatory condition for irrefutable traceability that overcomes the limitations of mere local authentication. The protocol captures operator face via device camera, extracts normalized 128-dimensional embedding $\mathbf{v}_{\text{capture}}$ using pre-trained neural networks (FaceNet, ArcFace), and loads stored template $\mathbf{v}_{\text{stored}}$ from the verified configuration. zkVM similarity computation receives public inputs (threshold $\tau$, user identifier) and private witness ($\mathbf{v}_{\text{capture}}$, $\mathbf{v}_{\text{stored}}$), generating proof $\pi_1$ asserting:
\begin{equation}
\exists \mathbf{v}_{\text{capture}}, \mathbf{v}_{\text{stored}} :
\frac{\mathbf{v}_{\text{capture}} \cdot \mathbf{v}_{\text{stored}}}{\|\mathbf{v}_{\text{capture}}\| \|\mathbf{v}_{\text{stored}}\|} \geq \tau
\end{equation}
without revealing embedding vectors, ensuring only authorized users activate agent services.

\textit{Phase 2---Agent Configuration Integrity.} Building upon operator verification, the protocol verifies that the agent's local configuration matches the blockchain-anchored canonical version. The on-chain $\mathsf{AgentID}$ stores $\text{profile\_hash} = H(\text{Config})$ in an Ethereum smart contract at storage slot $s$. The protocol invokes \texttt{eth\_getProof(contractAddr, [s], blockNumber)} to retrieve Merkle proofs linking the $\mathsf{AgentID}$ to blockchain state. zkVM verification receives public inputs ($r_{\text{storage}}$, expected profile hash) and private witness (local configuration $\text{Config}_{\text{local}}$, Merkle proof $\pi_{\text{merkle}}$), then proves:
\begin{equation}
H(\text{Config}_{\text{local}}) = \text{profile\_hash in } \mathsf{AgentID}
\end{equation}
Leveraging blockchain's tamper-proof storage establishes a cryptographic anchor for agent identity, generating proof $\pi_2$ that binds storage root and Merkle proof verification without revealing configuration contents.

\textit{Phase 3---Iterative Execution with Communication Provenance.} The system generates cryptographic proofs for complete execution traces across conversational turns. Each turn $t$ involves: (1)~input reception (user query $q_t$, previous state commitment $h_{t-1} = H(S_{t-1})$); (2)~LLM invocation (remote service response $a_t$); (3)~tool execution ($o_t = f_{a_t}(\text{params})$); (4)~state update ($S_t = (S_{t-1}, q_t, a_t, o_t)$). The integrated execution proof $\pi_t$ combines three verification stages within one zkVM execution:
\begin{enumerate}
\item \textit{Recursive Verification.} For $t > 1$, verify parent proof $\pi_{t-1}$ validity through zkVM's recursive verification capability---where Prove algorithm verifies another zkVM proof within its execution---preventing broken proof chains from extending. This constructs proof chains with cryptographically enforced sequential dependencies: proof $\pi_k$ embeds complete previous proof $\pi_{k-1}$ in public inputs $x_{pub,k}$, and constraint template includes recursive verification program $P_{\text{rec}}(vk, C_{P}, x_{pub,k-1}, y_{k-1}, \pi_{k-1})$. Successful verification of final proof $\pi_T$ implies the entire execution sequence $S_0 \to S_1 \to \cdots \to S_T$ validates in order with no permutation or substitution.

\item \textit{zkTLS Verification.} Execute TLS session validation within zkVM to ensure authenticity of external communications. zkVM operates in isolated environments that cannot verify external data sources---an adversary controlling execution environment can replace genuine API responses with fabricated data. We adopt zkTLS (Zero-Knowledge Transport Layer Security) to establish cryptographic provenance for HTTPS communications. The unified verification circuit validates: (a)~certificate validity (verify signature chain to trusted root CA, check certificate matches server identity); (b)~session key derivation (compute Pre-Master Secret, derive $K_{\text{session}}$ via PRF); (c)~data integrity (decrypt response $d = \text{D}_{K_{\text{session}}}(c)$, verify AEAD authentication tag); and (d)~commitment binding ($\text{commit}(d) = H(d)$ matches public input). The unified proof $\pi_t$ cryptographically attests: \textit{there exists a valid TLS session with server $S$ that decrypts to data $d$ (matching commitment $H(d)$), and program $P$ correctly processed $d$ to produce claimed output}---binding TLS communication provenance with computation correctness.

\item \textit{Agent Computation Verification.} Verify agent correctly executes tasks using verified responses: parse action, execute tool call, update state, compute state commitment $h_{S_t} = H(S_t)$.
\end{enumerate}

After $T$ conversational turns, final proof $\pi_T$ cryptographically attests to: (1)~initial authentication validity (Phase~1 configuration integrity, Phase~2 operator biometric verification); (2)~complete execution trace across all turns: $S_0 \to S_1 \to \cdots \to S_T$; (3)~verified provenance of all external service responses: $\{a_1, \ldots, a_T\}$ via zkTLS; (4)~correct agent computation for all tool executions: $\{o_1, \ldots, o_T\}$ following code $C_P$. 

\textbf{Layer 2: Permission Validation.} After establishing identity and execution integrity through Layer~1, the system validates authorization scope using Verifiable Credentials. The verifier validates agent's VC to confirm permission scope matches requested operation: (a)~signature verification using user's public key from User Identity Contract; (b)~credential validity period checking; (c)~permission scope matching against target task context; (d)~revocation status checking. This layer ensures authenticated agents operate within explicitly delegated permission boundaries, preventing privilege escalation.

\textbf{Authentication Output.} Upon successful completion, the protocol outputs: (1)~unified proof $\pi_T$ demonstrating agent configuration integrity, operator biometric verification, and complete execution trace; (2)~validated Verifiable Credential with verified permission scope. Verifiers (counterpart agents, service providers, regulators) independently verify by checking $\pi_T$ using zkVM's Verify algorithm against on-chain $\mathsf{AgentID}$ and validating VC signature and scope, establishing cryptographic guarantees for trustworthy agent interactions without centralized trust infrastructure.

\section{Implementation and Performance Evaluation}\label{sec:evaluation}

We evaluate the BAID system across two critical dimensions: (1)~on-chain identity management smart contracts performance focusing on gas consumption efficiency for on-chain identity management operations, and (2)~verifiable agent system performance focusing on zkVM-based authentication protocol efficiency. These evaluations address distinct but complementary aspects of the BAID architecture, demonstrating practical deployability across both on-chain and off-chain components.

\subsection{Identity Management Smart Contracts Performance}

To evaluate the gas consumption efficiency of the BAID on-chain identity management system, we deployed the complete smart contract architecture on the Ethereum Virtual Machine (EVM) testnet and conducted comprehensive performance testing using the Hardhat framework~\cite{hardhatgithub}, a development environment that provides deterministic gas metering consistent with mainnet behavior. We deployed the complete BAID contract suite including IdentityFactory, UserIdentityContract template, AgentIdentityContract template, and all zkKYC verification contracts. Gas measurements were collected using Hardhat's built-in gas reporter, which instruments the EVM to track computational costs for each transaction. The evaluation focused on gas consumption for core identity management operations, as gas costs directly impact the economic feasibility and scalability of blockchain-based identity systems.

Table~\ref{tab:gas-consumption} presents the gas consumption statistics for the four core BAID system operations. The measurements represent average values across multiple test executions, demonstrating consistent performance characteristics.

\begin{table}[htbp]
\caption{Gas Consumption for Core BAID System Operations}
\label{tab:gas-consumption}
\centering
\small
\begin{tabularx}{\columnwidth}{lX}
\toprule
\textbf{Operation} & \textbf{Gas Consumption} \\
\midrule
User Registration (\texttt{registerUser}) & 390,325 gas \\
Agent Registration (\texttt{registerAgent}) & 507,763 gas \\
Agent Deregistration (\texttt{deregisterAgent}) & 124,117 gas \\
Agent Update (\texttt{updateAgent}) & 128,837 gas \\
\bottomrule
\end{tabularx}
\end{table}

The experimental results demonstrate two fundamental characteristics of the BAID blockchain system's gas consumption profile:

\textbf{(1) Registration Cost Composition.} Registration operations exhibit high gas consumption driven by two factors: cryptographic verification overhead and storage allocation costs. User registration ($\sim$390K gas) performs zkKYC verification for privacy-preserving identity validation, consuming substantial gas for on-chain zero-knowledge proof verification. Agent registration ($\sim$508K gas) adds 30\% overhead through Clone-pattern contract instantiation and user-agent binding establishment. The fundamental storage cost asymmetry---SSTORE from zero to non-zero requires 20,000 gas per slot versus 5,000 gas for modifications---explains why registration operations dominate gas consumption.

\textbf{(2) Efficient State Modification.} Update ($\sim$129K gas) and deregistration ($\sim$124K gas) operations consume approximately 75\% less gas than registrations by avoiding expensive contract creation and leveraging existing storage slots. These operations achieve consistent performance through minimal state modifications and event emission, demonstrating the efficiency of the BAID contract architecture for routine identity management operations.

Overall, the BAID blockchain system achieves gas-efficient identity management through three design choices: zkKYC for privacy-preserving user registration, Clone pattern for agent contract deployment, and minimal storage footprint for affordable routine operations.

\subsection{Verifiable Agent System Performance}

To evaluate the performance and scalability of the BAID authentication protocol, we implemented a complete Verifiable Agent System using RISC Zero's zkVM framework~\cite{risc0github} and conducted comprehensive performance testing on the recursive proof generation pipeline. The testing environment consisted of a Rust-based zkVM execution framework implementing the complete BAID authentication pipeline described in Section~5.3, deployed on a standard development machine (Apple M1 chip, 16GB RAM) using RISC Zero zkVM version 1.0 with the composite proof system.

The implementation follows a fully recursive verification pipeline spanning five sequential phases: (1)~Phase~1 (Operator Biometric Authentication) establishes the baseline proof; (2)~Phase~2 (Agent Configuration Integrity) recursively verifies Phase~1; (3)~Phase~3 comprises three iterations of Iterative Task Execution, where Turn~1 recursively verifies Phases~1-2, Turn~2 recursively verifies Phases~1-2 and Turn~1, and Turn~3 recursively verifies the complete execution history from Phase~1 through Turn~2. The evaluation focused on execution efficiency, proof generation overhead, verification costs, and proof size characteristics across this multi-phase recursive authentication workflow, as these metrics directly impact the practical deployability and scalability of zkVM-based agent identity systems.

Table~\ref{tab:zkvm-performance} presents the performance characteristics for each phase of the BAID authentication protocol. The measurements represent average values across multiple test executions, demonstrating consistent performance patterns.

\begin{table}[htbp]
\caption{zkVM Performance Metrics for BAID Authentication Protocol}
\label{tab:zkvm-performance}
\centering
\small
\begin{tabularx}{\columnwidth}{lXXXXX}
\toprule
\textbf{Phase} & \textbf{Exec} & \textbf{Prove} & \textbf{Total} & \textbf{Proof} & \textbf{Verify} \\
 & \textbf{(ms)} & \textbf{(s)} & \textbf{(s)} & \textbf{(KB)} & \textbf{(ms)} \\
\midrule
Phase 1: Bio Auth & 9 & 15.00 & 15.01 & 238 & 14 \\
Phase 2: Config Veri & 13 & 31.35 & 31.36 & 488 & 28 \\
Phase 3 (Turn 1) & 17 & 40.14 & 40.16 & 737 & 43 \\
Phase 3 (Turn 2) & 19 & 37.68 & 37.70 & 987 & 74 \\
Phase 3 (Turn 3) & 15 & 38.29 & 38.31 & 1236 & 93 \\
\bottomrule
\end{tabularx}
\end{table}

\textbf{Proof Generation Overhead.} Proof generation dominates the computational cost, constituting 99.9\% of total processing time across all phases. For instance, Phase~1 requires only 9ms for program execution but 15.00s for proof generation, demonstrating that cryptographic proof construction overhead far exceeds the program execution itself. This asymmetry aligns with fundamental zkVM architecture where constraint satisfaction and polynomial commitment schemes require intensive computation, while program execution remains lightweight. Notably, proof generation time scales polylogarithmically with program size but exhibits near-constant complexity with respect to recursive depth---the 15-40 second generation latency remains consistent across phases despite accumulating recursive verification steps. This property makes the protocol suitable for high-security authentication scenarios where cryptographic guarantees justify computational overhead.

\textbf{Verification Overhead.} In contrast to proof generation, verification exhibits succinct characteristics with logarithmic complexity relative to program size. Verification time demonstrates linear growth with recursive depth---Phase~3 (Turn~1) requires 43ms, Phase~3 (Turn~2) (with one recursive verification) increases to 74ms, and Phase~3 (Turn~3) (with two recursive verifications) reaches 93ms. This linear scaling in recursive depth is the fundamental cost of ensuring execution sequence integrity: each additional turn cryptographically validates all previous computation, preventing proof chain manipulation or execution reordering attacks. The proof size similarly scales linearly with recursive depth (737KB for Turn~1, 987KB for Turn~2, 1236KB for Turn~3), reflecting the embedded verification history. Despite this linear growth, millisecond-level verification times remain practical for real-world deployment scenarios where cryptographic execution ordering guarantees are critical security requirements. This asymmetric cost structure---expensive proof generation concentrated at the prover, with efficient verification distributed to multiple verifiers---enables scalable multi-party verification in decentralized agent ecosystems without requiring verifiers to re-execute the complete agent workflow.

Overall, the zkVM evaluation demonstrates that the BAID authentication protocol achieves practical performance characteristics suitable for real-world deployment, with asymmetric costs favoring verifiers over provers in high-security authentication scenarios.

\section{Related Work}\label{sec:related}

Trustworthy AI agent collaboration requires three interconnected capabilities: standardized communication protocols, robust identity authentication, and operator accountability mechanisms. While existing research has advanced each dimension independently, a critical gap remains: no unified system simultaneously ensures agent identity integrity, operator accountability, and cross-protocol interoperability. We organize related work along these dimensions and identify limitations motivating our approach.

\textbf{Agent Communication Protocols.} The field of agent communication protocols has witnessed significant developments with several sophisticated frameworks emerging. Model Context Protocol (MCP) introduced by Anthropic~\cite{anthropic2024mcp,mcp2025specification,schmid2025mcp} establishes a comprehensive framework for contextual information exchange. Agent-to-Agent (A2A) Protocol~\cite{surapaneni2025announcing,a2a2025specification}, developed by Google, represents a breakthrough in opaque agent interoperability. Agent Communication Protocol (ACP)~\cite{LF2025ACPJoinsA2A} focuses on local network-prioritized communication. Additionally, several novel frameworks have been proposed: ANP (Agent Network Protocol)~\cite{lin2025anp} facilitates large-scale agent networking; LMOS (Language Model Operating System)~\cite{lmos2025} by Eclipse integrates LLMs into system-level operations; and the Agent Protocol~\cite{agentprotocol2025} offers an open-source, framework-agnostic standard based on OpenAPI v3 for managing agent lifecycles (start, stop, monitor). Despite their technical sophistication, these protocols share a fundamental limitation: the absence of a unified, trusted layer for agent registration and discovery, creating significant barriers to cross-protocol identity verification and trusted collaboration environments.

\textbf{Identity Authentication Frameworks.} Identity authentication frameworks, serving as fundamental infrastructure for trusted agent collaboration, have achieved significant advancements. Contemporary frameworks predominantly build upon OAuth 2.0~\cite{rfc6749} and OpenID Connect~\cite{openid}. While OAuth 2.0's Client Credentials flow theoretically enables agent authentication, its implementation reveals substantial limitations: the client\_id mechanism lacks cross-platform universality, and existing ID Tokens exclusively identify human users while Access Tokens fail to encapsulate agent capabilities. To address these limitations, South et al.~\cite{south2025} proposed the Authenticated Delegation framework, extending OAuth 2.0 and OpenID Connect with Agent-ID Tokens incorporating unique agent identifiers, capability manifests, and delegation metadata. However, this approach faces three key limitations: centralization dependencies limiting cross-domain agent identification; domain boundaries restricting system integration; and security vulnerabilities arising from centralized identity data management creating single points of failure. Moreover, these traditional frameworks fundamentally rely on key-based authentication, where identity is proven by possession of a private key rather than the integrity of the executing code. This creates a semantic gap for autonomous agents: proving ``who holds the key'' does not guarantee ``what code is running,'' leaving systems vulnerable to code substitution attacks where malicious agents generate valid signatures. While emerging Verifiable Computation paradigms, particularly Zero-Knowledge Virtual Machines (zkVM) and zkTLS, offer theoretical foundations for code-level authentication and communication provenance, they have yet to be integrated into a unified, practical agent identity framework that simultaneously ensures execution integrity and privacy.

\textbf{Agent Identity Management Systems.} Academic research has produced several groundbreaking approaches to agent identity management, which can be categorized into foundational identity infrastructures and decentralized trust frameworks. In the first category, Chan et al.~\cite{chan2024ids} pioneered the conceptual framework for AI system identity markers, creating criteria for accessibility and verifiability. This was further operationalized by Huang et al.\ with the Agent Name Service (ANS)~\cite{huang2025ans}, which implements unified registration and resolution through DNS-style naming and PKI. Advancing this further, Raskar et al.~\cite{raskar2025upgrade} proposed hybrid indexing approaches that integrate with existing DNS/PKI, leading to the NANDA (Beyond DNS)~\cite{nanda2025beyonddns} framework. NANDA combines a global Core Index with a verifiable AgentFact framework to enable trusted discovery.

Complementing these foundational layers, emerging decentralized trust frameworks address the specifics of agent autonomy and ethics. The LOKA protocol~\cite{ranjan2025loka} serves as a decentralized identity framework specifically targeting accountability and ethical consistency, implementing a blockchain-based trust layer to govern agent interactions. Similarly, ERC-8004~\cite{erc8004} introduces ``Trustless Agents'' via an on-chain validation registry, allowing agents to submit verifiable proofs of their work. However, while these systems successfully implement technical registration and discovery mechanisms, they predominantly focus on ``agent-to-system'' trust---verifying that the agent is a technically valid and recognizable entity within the network. They largely overlook the critical ``human-to-agent'' liability binding---cryptographically anchoring the agent's actions to a specific natural person to ensure legal accountability. This omission makes it difficult to establish the comprehensive responsibility chains necessary for autonomous deployment.

To address these critical gaps, we propose the BAID (Binding Agent ID) framework, which implements an innovative dual-mechanism approach combining local binding and on-chain identity verification. This comprehensive solution not only bridges the technical gaps in responsibility binding but also establishes a robust foundation for secure AI agent operations in complex social collaboration environments. Our framework represents a significant advancement in reconciling technical identification requirements with legal traceability demands, offering a practical path forward for responsible AI agent deployment.

\section{Conclusion}\label{sec:conclusion}

This paper presents BAID (Binding Agent ID), a comprehensive identity infrastructure addressing the fundamental challenge of autonomous AI agents: the absence of traceable and accountable responsibility chains. BAID implements the Binding Agent Model through three integrated mechanisms: (1)~local binding via biometric authentication ensures agents serve only their designated owners, preventing command injection attacks; (2)~on-chain identity registration via blockchain smart contracts enables trusted agent discovery and establishes publicly auditable user-agent binding declarations; (3)~zkVM-based authentication protocol provides cryptographic guarantees for operator identity verification, agent configuration integrity, and complete execution provenance through recursive proof composition. This unified framework addresses the five critical security requirements identified in cross-organizational agent interactions: user identity verification, trusted discovery, mutual authentication, permission control, and accountability tracking. Our implementation and evaluation demonstrate practical deployability: blockchain identity management achieves viable gas consumption for real-world Ethereum deployment, while zkVM authentication exhibits millisecond-level verification latency suitable for security-prioritized scenarios despite proof generation overhead. Future work should address performance optimization to enable broader adoption beyond specialized high-security use cases.

\bibliographystyle{cas-model2-names}
\bibliography{cas-refs}

\appendix

\section{zkVM-based Authentication Protocol Technical Specifications}

This appendix provides comprehensive technical specifications for the zkVM-based verifiable computation protocol that establishes cryptographic guarantees for agent identity integrity, operator authentication, and execution provenance. The protocol addresses the fundamental challenge in autonomous AI systems---ensuring the correct agent is used by the correct user in the correct manner---thereby constructing an end-to-end trust chain from identity validation to execution auditing.

\subsection{Limitations of Key-Based Authentication}

Traditional software authentication relies on digital signature schemes where possession of private key $sk$ enables signing operations that prove ownership. While effective for human-controlled systems, this paradigm exhibits a critical vulnerability when applied to autonomous agents: key compromise enables arbitrary code substitution.

An adversary obtaining the agent's signing key $sk$ can: (1) execute malicious code $P_{\text{malicious}}$ instead of legitimate program $P_{\text{legitimate}}$; (2) generate valid signatures $\sigma = \text{Sign}_{sk}(\text{output})$ over malicious outputs; (3) present outputs as originating from the legitimate agent. Verification $\text{Verify}_{pk}(\text{output}, \sigma)$ succeeds, yet computation violates intended behavior, breaking the assumption that cryptographic authenticity implies behavioral integrity.

Unlike human users, software agents lack physical embodiment amenable to conventional biometric authentication. The agent's identity is inherently tied to its computational behavior---the code it executes. Traditional authentication verifies the credential holder, not the executing code, creating a semantic gap between ``who holds the key'' and ``what code is running.''

\subsection{Code-Level Authentication}

We introduce code-level authentication, a cryptographic identity model where the program binary serves as the agent's unique identity marker, shifting authentication from key-based ownership proofs to code-based execution verification.

Based on the zkVM five-tuple from Section 2:
\[\Pi = (\text{Setup}, \text{CommitProg}, \text{Compile}, \text{Prove}, \text{Verify}),\]

\textbf{Definition 1 (Execution Integrity Relation).} The protocol proves:
\begin{equation}
\begin{split}
\mathcal{R}_{\text{exec}} = \{(P, x_{pub}, x_{prv}, y) : P(x_{pub}, x_{prv}) = y \\ \land \text{CommitProg}(P) = C_P\}
\end{split}
\end{equation}
where $P$ is the agent program, $x_{pub}$ are public inputs,
$x_{prv}$ are private inputs, and $y$ is the public output.

\textbf{Property 1 (Code Substitution Resistance).} Any modification to $P$ yields
$\text{CommitProg}(P') \neq C_P$, causing proof construction to fail during zkVM constraint verification.

\textbf{Security Framework.} The security framework of zkVM encompasses three fundamental guarantees:
\begin{itemize}
    \item \textbf{Completeness:} For any valid statement, an honest prover can consistently generate proofs that are accepted by the verification algorithm with probability 1.

    \item \textbf{Soundness:} For any invalid statement, no computationally bounded adversarial prover can generate an accepting proof, except with negligible probability in the security parameter. The verification algorithm systematically rejects proofs of false statements.

    \item \textbf{Zero-Knowledge:} For any valid statement, the proof reveals no information beyond its validity. Specifically, the verifier gains no computational advantage in deriving any additional knowledge about the private inputs or intermediate computation states.
\end{itemize}

\subsection{zkVM-Based Agent Authentication Protocol Workflow}

The protocol workflow comprises four phases:

\textbf{Initialization:} $(pp, vk) \leftarrow \text{Setup}(1^\lambda, param)$ generates system parameters.

\textbf{Registration:} Compile agent code to obtain
\[(\text{ArithDesc}, \text{MapIO}) \leftarrow \text{Compile}(pp, P);\]
generate commitment
\[C_P \leftarrow \text{CommitProg}(P);\]
construct $\mathsf{AgentID}$ following Equation~\eqref{eq:baid-structure}; anchor $\mathsf{AgentID}$  to blockchain smart contract.

\textbf{Execution:} For each step $k$, generate proof
$\pi_k \leftarrow \text{Prove}(pp, C_P, \text{ArithDesc}, \text{MapIO}, x_{pub}, x_{prv}, y)$
where execution trace $\text{Trace}(P, x_{pub}, x_{prv})$ is validated through constraint templates.

\textbf{Verification:} Compute $b \leftarrow \text{Verify}(vk, C_P, x_{pub}, y, \pi)$,
confirming proof validity and code commitment $C_P$ matches the registered $\mathsf{AgentID}$ value.

\subsection{Execution Continuity Across Conversational Turns}

\subsubsection{Problem Statement}

LLM-based agents operate through conversational state transitions involving user input reception,
remote LLM API invocation, tool execution, and state updates.
A fundamental zkVM constraint is that while individual proofs can certify that a computation
correctly transforms a given input to its output, they do not verify the provenance of inputs.

Given execution sequence:
\begin{equation}
S_0 \xrightarrow{a_1, f_1} S_1 \xrightarrow{a_2, f_2} S_2 \xrightarrow{a_3, f_3} \cdots \xrightarrow{a_n, f_n} S_n
\end{equation}
where $a_i$ are actions and $f_i$ are transition functions,
standard zkVM can generate independent proofs $\{\pi_1, \pi_2, \ldots, \pi_n\}$.
Each $\pi_i$ proves: ``given some value claimed to be $S_{i-1}$, the computation
$f_i(S_{i-1}, a_i)$ correctly produces $S_i$.''
However, $\pi_i$ does not verify that this $S_{i-1}$ is actually the output from $\pi_{i-1}$.

Without cryptographic binding between proofs, a verifier receiving $\{\pi_1, \pi_2\}$ cannot distinguish:
valid sequence $S_0 \xrightarrow{\pi_1} S_1 \xrightarrow{\pi_2} S_2$;
invalid permutation $S_0 \xrightarrow{\pi_2} S_2' \xrightarrow{\pi_1} S_1'$;
or invalid replay $S_0 \xrightarrow{\pi_1} S_1 \xrightarrow{\pi_1} S_1$.
Input provenance independence breaks execution ordering guarantees.

\subsubsection{Recursive Proof Algorithm}

We leverage zkVM's recursive verification capability---where the zkVM's Prove algorithm
can verify another zkVM proof within its execution---to construct proof chains with cryptographically
enforced sequential dependencies.

\textbf{Recursive Verification Program.} Define special program $P_{\text{rec}}$ with functionality:
\begin{equation}
\begin{split}
P_{\text{rec}}(vk, C_{P}, x_{pub,k-1}, y_{k-1}, \pi_{k-1}) \to \{0,1\}
\end{split}
\end{equation}

This program executes
\[\text{Verify}(vk, C_{P}, x_{pub,k-1}, y_{k-1}, \pi_{k-1})\]
within the zkVM, outputting the verification result.

\textbf{Proof Chain Construction.} For execution sequence $S_0 \to S_1 \to \cdots \to S_T$, step $k$ generates compound proof:
\begin{equation}
\begin{split}
\pi_k \leftarrow \text{Prove}(pp, C_P, \text{ArithDesc}_k, \\
\text{MapIO}_k, x_{pub,k}, x_{prv,k}, y_k)
\end{split}
\end{equation}

where:
\begin{itemize}
\item $x_{pub,k}$ contains complete previous proof $\pi_{k-1}$, program commitment $C_P$, verification key $vk$, previous output $y_{k-1}$, and previous claim $x_{pub,k-1}$
\item $x_{prv,k}$ contains current witness data
\item Constraint template $\text{ArithDesc}_k$ includes call to $P_{\text{rec}}(vk, C_{P}, x_{pub,k-1}, y_{k-1}, \pi_{k-1})$
\end{itemize}

\textbf{Security Properties.}

\textit{Property 2 (Unforgeability of Execution Order).} For proof chain $\{\pi_1, \pi_2, \ldots, \pi_T\}$, by induction on $k$:
\begin{itemize}
\item Base case ($k=1$):
\begin{align*}
\pi_1 \leftarrow \text{Prove}(pp, C_P, \text{ArithDesc}_1, \text{MapIO}_1, \\
x_{pub,1}, x_{prv,1}, y_1)
\end{align*}
where $x_{pub,1}$ contains no dependencies (initial step),
$\text{Verify}(vk, C_P, x_{pub,1}, y_1, \pi_1) = 1$ validates directly.

\item Inductive step: Assume for $k-1$,
\[\text{Verify}(vk, C_P, x_{pub,k-1}, y_{k-1}, \pi_{k-1}) = 1.\]
For step $k$, recursive verification program $P_{\text{rec}}$ executes within zkVM: (1) Extract previous proof $\pi_{k-1}$ from $x_{pub,k}$; (2) Verify parent proof:
\[b_{k-1} \leftarrow \text{Verify}(vk, C_{P_{k-1}}, x_{pub,k-1}, y_{k-1}, \pi_{k-1});\]
(3) If $b_{k-1} = 0$, $P_{\text{rec}}$ fails, cannot generate $\pi_k$; (4) If $b_{k-1} = 1$, generate integrated proof $\pi_k$, whose verification implies validity of $\pi_{k-1}$.
\end{itemize}

Therefore:
\begin{equation*}
\begin{split}
\text{Verify}(vk, C_P, x_{pub,k}, y_k, \pi_k) = 1 \\
\implies \text{Verify}(vk, C_P, x_{pub,k-1}, y_{k-1}, \pi_{k-1}) = 1.
\end{split}
\end{equation*}

\textit{Corollary.} Successful verification of final proof $\pi_T$ implies the entire execution sequence $S_0 \to S_1 \to \cdots \to S_T$ validates in order, with no permutation or substitution.

\textit{Property 3 (Fail-Stop Security).} If any intermediate step $t < T$ verification fails, i.e.,
\[\text{Verify}(vk, C_P, x_{pub,t}, y_t, \pi_t) = 0,\]
then: (1) Step $t+1$'s recursive verification program $P_{\text{rec}}$ detects failure
when verifying $\pi_t$ (condition C2 unsatisfied); (2) $P_{\text{rec}}$ execution fails,
triggering constraint violation; (3) zkVM cannot generate valid proof $\pi_{t+1}$ for step $t+1$;
(4) All subsequent steps $t+2, t+3, \ldots, T$ likewise cannot generate proofs.

Broken proof chains cannot be extended. Attackers cannot ``skip over'' failed steps---each
proof $\pi_{t+1}$ cryptographically embeds verification of $\pi_t$ within its recursive proof;
missing or invalid $\pi_t$ causes verification failure inside $P_{\text{rec}}$.

\textit{Property 4 (Compact Verification).} External verifiers need only check the final proof $\pi_T$:
\begin{itemize}
\item Verification complexity:
$\mathcal{O}(\log T) \cdot |\text{ArithDesc}|$\\
(STARK verification scales logarithmically;\\
$|\text{ArithDesc}|$ is constant)
\item Proof size: $|\pi_T| = \mathcal{O}(1)$ (constant size ~150-250 KB,
independent of execution length $T$)
\item Recursive embedding: All historical validations $\{\pi_1, \ldots, \pi_{T-1}\}$
are recursively verified and embedded into $\pi_T$ through the public input $x_{pub,T}$
\end{itemize}

\subsection{External Communication Provenance}

\subsubsection{Problem Statement}

zkVM operates within an isolated execution environment that cannot directly access network resources or verify external data sources. When an agent invokes external tools (e.g., web search APIs, databases), remote LLM services (e.g., OpenAI GPT-4, Anthropic Claude), or other agents (in multi-agent systems), \textbf{the zkVM cannot distinguish authentic API responses from fabricated data provided by the external environment}.

An adversary controlling the execution environment can: (1) Intercept the agent's intended API call (e.g., query to LLM service); (2) Replace the genuine response with attacker-controlled content; (3) Feed the fabricated data to the zkVM's Prove algorithm. The zkVM generates a valid proof $\pi$ over the tampered input, yet the computation violates the agent's declared behavior. The proof attests to computational correctness, not data provenance.

\subsubsection{zkTLS Integration for Verifiable HTTPS Communications}

We adopt \textbf{zkTLS (Zero-Knowledge Transport Layer Security)} to establish cryptographic provenance for external communications. zkTLS extends TLS with zero-knowledge proofs, enabling an agent to prove properties about HTTPS sessions (e.g., ``I received data $d$ from server $S$'') without revealing the session contents to third parties.

\textbf{Standard TLS Workflow:}
\begin{enumerate}
\item Handshake: Client and server exchange certificates, negotiate cipher suite, derive session keys
\item Data Transfer: Application data is encrypted with session keys, authenticated with AEAD (e.g., AES-GCM)
\end{enumerate}

\textbf{zkTLS Extension:}
\begin{enumerate}
\item Transcript Recording: The client (agent) records the complete TLS transcript, including handshake messages and encrypted data
\item Zero-Knowledge Proof Generation: The client generates a proof asserting that:
\begin{itemize}
\item[(a)] The TLS handshake is valid (certificate chain verified, session keys correctly derived)
\item[(b)] Specific data fields $d$ were present in the server's response
\item[(c)] The proof reveals no information about the full response content or session keys
\end{itemize}
\end{enumerate}

\textbf{Protocol Integration with zkVM.} For each external communication at step $k$:

\textbf{Step1: TLS Session Execution (External Environment).} The execution environment:
(1) Initiates TLS connection to target server $S$ (e.g., \texttt{api.openai.com});
(2) Sends API request (e.g., LLM prompt);
(3) Receives encrypted response $c = \text{E}_{K_{\text{session}}}(d)$;
(4) Records full TLS transcript:
\[T = (\text{ClientHello}, \text{ServerHello}, \text{Certificate}, c, \ldots).\]

\textbf{Step2: Integrated zkVM Execution (Inside zkVM).} The zkVM's Prove algorithm receives:
\begin{itemize}
\item Public inputs: server identity $S$, data commitment $\text{commit}(d)$
\item Private witness: TLS transcript $T$, session keys $K_{\text{session}}$
\end{itemize}

The zkVM executes a unified verification and computation circuit that:

\textbf{(a) Validates TLS Communication Provenance:}
\begin{enumerate}
\item \textbf{Certificate Validity:} Parse X.509 certificate from $T$, verify signature chain up to trusted root CA, and check certificate matches server identity $S$ (DNS name in Subject Alternative Name field)

\item \textbf{Session Key Derivation:} Extract client/server random nonces from ClientHello/ServerHello, compute Pre-Master Secret (from RSA/ECDHE key exchange), and derive session keys via PRF:
\[K_{\text{session}} = \text{PRF}(\text{PMS}, \text{nonces}, \text{``key expansion''})\]

\item \textbf{Data Integrity:} Decrypt response $d = \text{D}_{K_{\text{session}}}(c)$, and verify AEAD authentication tag (e.g., GCM tag) to ensure $d$ was not tampered

\item \textbf{Commitment Binding:} Compute $\text{commit}(d) = H(d)$ and check commitment matches the public input
\end{enumerate}

\textbf{(b) Proves Agent Computation with Verified Data:} The zkVM proves that the agent's computation correctly uses the verified data $d$: parsing the data, performing business logic operations, and generating computational outputs.

\textbf{Output:} The zkVM generates a single unified proof $\pi_{k}$ that cryptographically attests: \textit{there exists a valid TLS session with server $S$ that decrypts to data $d$ (matching the public commitment $H(d)$), and the agent program $P$ correctly processed $d$ to produce the claimed output}. This proof binds TLS communication provenance with computation correctness without revealing the private TLS transcript, session keys, or decrypted data.

\textbf{Security Guarantee:} An attacker cannot: (1) substitute $d$ with fabricated data $d'$ (breaks hash binding $H(d) \neq H(d')$); (2) claim $d$ came from different server $S'$ (breaks TLS certificate verification); (3) modify $d$ after TLS decryption (breaks AEAD authentication tag); or (4) manipulate computation with incorrect data (breaks unified proof integrity).

\subsection{Three-Phase Verification Pipeline}

The BAID protocol orchestrates agent authentication through three sequential phases,
each enforcing specific security invariants:
\begin{equation*}
\begin{split}
\text{Phase 1 (Operator)} \to \text{Phase 2 (Config)} \\
\to \text{Phase 3 (Execution)}
\end{split}
\end{equation*}

Each phase generates a proof $\pi_i$ that feeds into the next phase via recursive proof algorithm,
ensuring end-to-end verification.

\subsubsection{Phase 1: Operator Biometric Authentication}

\textbf{Objective:} Verify the operator's identity without revealing biometric templates, ensuring only authorized users can activate the agent.

\textbf{Privacy-Preserving Facial Recognition:}
\begin{enumerate}
\item Feature Extraction: Capture operator's face via device camera;
extract 128-dimensional embedding $\mathbf{v}_{\text{capture}} \in \mathbb{R}^{128}$
using a pre-trained neural network (e.g., FaceNet, ArcFace);
normalize: $\mathbf{v}_{\text{capture}} \leftarrow \frac{\mathbf{v}_{\text{capture}}}{\|\mathbf{v}_{\text{capture}}\|}$.

\item Reference Template Retrieval: Load stored template $\mathbf{v}_{\text{stored}} \in \mathbb{R}^{128}$
from the agent's security configurations file.

\item zkVM Similarity Computation: Public inputs (similarity threshold $\tau$, user identifier $\mathsf{UserID}$);
Private witness ($\mathbf{v}_{\text{capture}}$, $\mathbf{v}_{\text{stored}}$).
Verification logic computes cosine similarity and performs threshold comparison.

\item Output: Proof $\pi_1$ where the zkVM generates proof asserting:
\begin{equation}
\exists \mathbf{v}_{\text{capture}}, \mathbf{v}_{\text{stored}} :
\frac{\mathbf{v}_{\text{capture}} \cdot \mathbf{v}_{\text{stored}}}{\|\mathbf{v}_{\text{capture}}\| \|\mathbf{v}_{\text{stored}}\|} \geq \tau
\end{equation}
Public output indicates authentication result (true/false), without revealing the actual embedding vectors.
The proof $\pi_1$ establishes operator authentication as the foundation for subsequent phases.
\end{enumerate}

\subsubsection{Phase 2: Agent Configuration Integrity}

\textbf{Objective:} Prove that the agent's local configuration matches
the blockchain-anchored canonical version.

\textbf{Protocol Steps:}
\begin{enumerate}
\item On-Chain Anchor: The Agent Identifier, defined in Equation~\eqref{eq:baid-structure}, is stored in an Ethereum smart contract, where $C_P = \text{CommitProg}(P)$ is the program commitment and $H(\text{profile})$ represents the hash of the agent's security configurations file (containing biometric features, behavioral rules, and security policies). The $\mathsf{AgentID}$ is anchored at a specific storage slot $s$ within the contract's state trie.

\item Merkle Proof Retrieval: The protocol invokes the Ethereum JSON-RPC method
\texttt{eth\_getProof( \\ contractAddr, [storageSlot], blockNumber)} to retrieve cryptographic proofs
linking the on-chain $\mathsf{AgentID}$ to the blockchain state. The method returns three components:
\begin{itemize}
\item Account proof: Merkle branch proving the contract account's existence in the state trie
\item Storage proof: Merkle branch proving that $(s, \mathsf{AgentID})$ is stored in the contract's storage trie
\item Storage root: $r_{\text{storage}}$ from the account state, serving as the verification anchor
\end{itemize}

\item zkVM Verification: Public inputs ($r_{\text{storage}}$, expected profile hash from $\mathsf{AgentID}$);
Private witness (local security configurations file $\text{Config}_{\text{local}}$, Merkle proof $\pi_{\text{merkle}}$).
Verification logic computes $H(\text{Config}_{\text{local}})$, verifies it matches $\text{profile\_hash}$ in $\mathsf{AgentID}$,
reconstructs storage root from leaf, and checks consistency.

\item Output: Proof $\pi_2$ asserting that the local configuration matches the blockchain-anchored version:
\begin{equation*}
H(\text{Config}_{\text{local}}) = \text{profile\_hash in $\mathsf{AgentID}$}
\end{equation*}
The proof binds the storage root $r_{\text{storage}}$ and Merkle proof verification without revealing the configuration contents.
The proof $\pi_2$ depends on $\pi_1$ through recursive proof algorithm.
\end{enumerate}

\subsubsection{Phase 3: Iterative Task Execution}

\textbf{Objective:} Prove correct execution of agent tasks across multiple conversational turns
$t = 1, 2, \ldots, T$, with verified provenance for all external communications.

\textbf{Execution Model per Turn $t$:} Each conversational turn $t$ involves:
(1) Input Reception (user query $q_t$ or tool response $r_{t-1}$, previous state commitment $h_{t-1} = H(S_{t-1})$);
(2) LLM Invocation (send prompt to remote LLM service, receive response $a_t$);
(3) Tool Execution (execute tool function: $o_t = f_{a_t}(\text{params})$);
(4) State Update (compute new state: $S_t = (S_{t-1}, q_t, a_t, o_t)$).

\textbf{Integrated Execution Proof per Turn $t$:} For each conversational turn $t$,
generate a single execution proof $\pi_t$ that integrates zkTLS verification, agent computation,
and recursive proof algorithm within one zkVM execution. The proof generation proceeds in three stages:

1. Recursive Verification and Dependency Checking:
For $t > 1$, first verify parent proof $\pi_{t-1}$ validity through zkVM's recursive verification.
This ensures the current turn builds on a verified historical foundation,
preventing broken proof chains from extending.

2. zkTLS Verification (Communication Provenance):
Execute TLS session validation within zkVM to ensure authenticity of responses from LLM services and remote third-party APIs:
verify certificate chain, session key derivation, AEAD authentication,
and compute data commitment $h_{a_t} = H(a_t)$.

3. Agent Computation Verification (Execution Correctness):
Verify the agent correctly executes tasks using verified responses from external services:
parse action, execute tool call, update state, and compute state commitment $h_{S_t} = H(S_t)$.

4. Final Proof after $T$ Turns: After $T$ conversational turns,
the final proof $\pi_T$ cryptographically attests to:
(1) Initial authentication validity (Phase 1 configuration integrity and Phase 2 operator biometric verification);
(2) Complete execution trace across all turns: $S_0 \to S_1 \to \cdots \to S_T$;
(3) Verified provenance of all external service responses: $\{a_1, \ldots, a_T\}$ via zkTLS;
(4) Correct agent computation for all tool executions: $\{o_1, \ldots, o_T\}$ following code $C_P$.

External verifiers (eg. counterpart agents, service providers, or regulators) only need to:
(1) Receive the final proof $\pi_T$;
(2) Verify using zkVM's Verify algorithm: $b \leftarrow \text{Verify}(vk, C_P, x_{pub}, y_T, \pi_T)$;
(3) Check the public outputs match expected values.

\section{Security Analysis}
\label{sec:security_analysis}

This section analyzes the security properties of the BAID framework, defining the threat model and demonstrating how the proposed protocols defend against critical attacks targeting autonomous agent systems.

\subsection{Threat Model}

We consider an adversary $\mathcal{A}$ with the following capabilities:
\begin{itemize}
    \item \textbf{Key Compromise:} $\mathcal{A}$ may obtain the private signing keys of legitimate users or agents.
    \item \textbf{Network Control:} $\mathcal{A}$ can intercept, modify, replay, or reorder network messages between agents, users, and external services.
    \item \textbf{Malicious Execution Environment:} $\mathcal{A}$ may control the computational environment where the agent executes, enabling manipulation of I/O or substitution of program binaries.
    \item \textbf{Data Fabrication:} $\mathcal{A}$ can forge responses from external APIs (e.g., LLMs, web services) to mislead agent execution.
\end{itemize}

We assume the underlying cryptographic primitives (zkVM proof system, hash functions, digital signatures, TLS) are secure.

\subsection{Security Properties and Defenses}

\subsubsection{Defense Against Code Substitution Attacks}
\textbf{Attack:} An adversary with a compromised agent private key replaces the legitimate agent program $P$ with malicious code $P'$ to execute unauthorized actions while signing them as the legitimate agent.

\textbf{Defense:} BAID implements \textit{Code-Level Authentication}. The agent's identity is cryptographically bound to the program commitment $C_P = \text{CommitProg}(P)$. Any modification to the code results in $C_{P'} \neq C_P$. The zkVM proof generation requires the execution trace to satisfy constraints derived from $C_P$~\cite{risc0github}. Therefore, $\mathcal{A}$ cannot generate a valid proof $\pi$ for $P'$ that verifies against the legitimate $\mathsf{AgentID}$ (which contains $C_P$), effectively preventing code substitution even if keys are compromised.

\subsubsection{Defense Against Replay and Reordering Attacks}
\textbf{Attack:} $\mathcal{A}$ intercepts valid proofs $\{\pi_1, \pi_2, \ldots\}$ from a session and replays them or reorders them (e.g., skipping a payment authorization step) to manipulate the agent's state transition.

\textbf{Defense:} BAID employs \textit{Recursive Verification} to enforce \textit{Execution Continuity}. Each proof $\pi_t$ cryptographically embeds the verification of the previous proof $\pi_{t-1}$ as a public input. This creates an immutable, ordered hash chain $S_0 \to S_1 \to \cdots \to S_t$. A replayed or reordered proof will fail the recursive verification check $P_{\text{rec}}$ inside the zkVM~\cite{risc0github}, as the input state commitment $h_{t-1}$ will not match the output of the previous valid step.

\subsubsection{Defense Against Data Fabrication (Man-in-the-Middle)}
\textbf{Attack:} $\mathcal{A}$ controls the network or execution environment and feeds fabricated LLM responses or market data to the agent to trigger incorrect decisions.

\textbf{Defense:} BAID integrates \textit{zkTLS} to ensure \textit{Data Provenance}. The zkVM circuit verifies the TLS handshake and server signatures within the zero-knowledge environment. It proves that the data $d$ processed by the agent originated from a specific, authenticated server $S$ (e.g., \texttt{api.openai.com}) and has not been tampered with~\cite{Zhang2020DECO}. $\mathcal{A}$ cannot forge a valid TLS transcript and corresponding zkVM proof without breaking the underlying TLS cryptography.

\subsubsection{Defense Against Sybil and Impersonation Attacks}
\textbf{Attack:} $\mathcal{A}$ creates multiple fake agent identities to flood the network or impersonates a reputable agent.

\textbf{Defense:} BAID relies on \textit{On-Chain Identity Management} and \textit{zkKYC}. Agent identities must be bound to a valid User Identity Contract, which requires zkKYC verification of the legal entity~\cite{pauwels2021zkkyc}. This imposes a real-world cost and accountability on identity creation, mitigating Sybil attacks. Furthermore, the on-chain binding of $\mathsf{AgentID}$ to $C_P$ and $\mathsf{UserID}$ ensures that any impersonation attempt is detectable by verifying the agent's proofs against the immutable blockchain registry~\cite{chan2024ids}.

\end{document}